%% file: ms.tex
\documentclass[letterpaper,twocolumn,10pt]{article}
\usepackage{usenix2019}
\usepackage{epsfig}
\include{mathpmtc} 
\usepackage{latexsym}
\usepackage{url}
\usepackage{amsmath}
\usepackage{hyperref}
\usepackage{wrapfig}  
\usepackage[normalem]{ulem}
\usepackage{enumerate}
\usepackage{xspace}
\usepackage{fancyhdr}
\usepackage{proof}
\usepackage{setspace}
\usepackage{graphicx}
\usepackage{array}
\usepackage{caption}
\usepackage{subcaption}
\usepackage{algorithm}
\usepackage[noend]{algpseudocode}
\usepackage{placeins}
\usepackage{pifont}
\usepackage{float}
\usepackage{adjustbox}
\usepackage{cancel}
\usepackage{fullpage}
\usepackage{multirow}
\usepackage{booktabs}
\usepackage{makecell}
\usepackage{pgffor}
\usepackage{afterpage}
\usepackage{xcolor}
\usepackage[compact]{titlesec}

\usepackage{listings}
\usepackage{tcolorbox}
\usepackage{amsthm}
\usepackage{subcaption}
\usepackage{url}
\usepackage{balance}

\usepackage{lstlang}
\usepackage{local}

\newif\ifdraft\drafttrue

\newcommand\sys{Contra\xspace}

\newcommand{\IE}{\emph{i.e.}}
\newcommand{\EG}{\emph{e.g.}}
\newcommand{\smsg}[1]{\mbox{\small \sc #1}}

\definecolor{light-gray}{gray}{0.8}
\definecolor{amethyst}{rgb}{0.6, 0.4, 0.8}
\definecolor{mapleleafblue}{rgb}{0,0.1,0.5}

\newtheorem*{attempt-versionedprobes}{\hspace{-0pt}Refinement (Versioned probes)}
\newtheorem*{attempt-limitedprobefreq}{\hspace{-0pt}Refinement (Limited probe frequency)}
\newtheorem*{attempt-expiration}{\hspace{-0pt}Refinement (Expiration)}
\newtheorem*{attempt-flowlet}{\hspace{-0pt}Refinement (Policy-aware flowlet switching)}
\newtheorem*{final}{\hspace{-0pt}Final solution}

\newtcbox{\codebox}{nobeforeafter,colframe=light-gray,colback=light-gray!10!white,boxrule=0.5pt,arc=4pt,
	boxsep=0pt,left=1pt,right=1pt,top=2pt,bottom=2pt,tcbox raise base}
\newenvironment{packeditemize}{
\begin{itemize}
	\setlength{\itemsep}{0.3pt}
	\setlength{\parskip}{2pt}
	\setlength{\parsep}{0pt}
}{\end{itemize}}

\newcounter{mylineno}

\begin{document}

\title{Contra: A Programmable System for Performance-aware Routing \vspace{4mm}}

\author{{\rm Kuo-Feng Hsu$^\dag$,} {\rm Ryan Beckett$^\star$}, {\rm Ang Chen$^\dag$}, {\rm Jennifer Rexford$^\ddag$}, {\rm Praveen Tammana$^\ddag$}, {\rm David Walker$^\ddag$}\\[0.2mm] 
{\normalsize\rm $^\dag$Rice University, $^\star$Microsoft Research, $^\ddag$Princeton University}}

%%\vspace{2mm}

\date{}

\maketitle

\subsection*{Abstract}
\input{abstract}

\input{introduction}
\input{language}
\input{overview2}
\input{compilation}
\input{dynamic}
\input{evaluation}
\input{relatedwork}
\input{conclusion}

\bibliographystyle{abbrv}
\bibliography{paper}

\balance

%%%\newpage
%%%\clearpage
%%%\appendix
%%%\input{appendix}

\end{document}

%% file: abstract.tex
\noindent
We present \sys, a system for performance-aware routing that can adapt to traffic changes at hardware speeds. 
While existing work has developed point solutions for performance-aware routing on a fixed topology (e.g., a Fattree) 
with a fixed routing policy (e.g., use least utilized paths), \sys can be configured to operate seamlessly over any network 
topology and a wide variety of sophisticated routing policies. 
Users of \sys write network-wide policies that rank network paths given their current performance. 
A compiler then analyzes such policies in conjunction with the network topology and decomposes them into switch-local P4 programs, 
which collectively implement a new, specialized distance-vector protocol. This protocol generates compact probes that traverse the network, 
gathering path metrics to optimize for the user policy dynamically. Switches respond to changing network conditions at hardware speeds 
by routing flowlets along the best policy-compliant paths. Our experiments show that \sys scales to large networks, 
and that in terms of flow completion times, it is competitive with hand-crafted systems that have been customized for specific topologies and policies.

%% file: introduction.tex
\section{Introduction}
\label{sec:intro}

Configuring a network to achieve a diverse range of objectives, such as 
routing constraints (\EG, traffic should go through a series of middleboxes),  
and traffic engineering (\EG, minimize latency and maximize throughput), is a challenging task. To handle
this complexity, one approach has been to use SDN solutions, which have a centralized point for management~\cite{hong-2013-swan,jain-2014-b4}. 
However, centralized controllers are inherently too slow to respond to fine-grained traffic changes, 
such as short traffic bursts. 
In fact, even the software control planes locally on the switches are often limited in their ability to make forwarding decisions 
fast enough.

Recent work has developed load balancing mechanisms that operate entirely in the data plane to enable real-time 
adaptation~\cite{alizadeh-2014-conga,katta-2016-hula}. 
By making use of fine-grained performance information on hardware timescales, these systems can deliver considerable performance benefits 
over static load balancing mechanisms like ECMP. 
Unfortunately, existing systems, such as Conga~\cite{alizadeh-2014-conga} and Hula~\cite{katta-2016-hula}, 
are point solutions that only work under very specific assumptions about the network topology, routing constraints, and performance objectives---they
only support a ``least utilized shortest path'' policy on a data center topology. 
Further, it is not obvious how to extend them to work in other kinds of networks (\EG, a WAN) or with more 
sophisticated policies. 

In this paper, we describe \sys, a general and programmable system for performance-aware routing. 
Network operators configure \sys by describing the
network topology as well as a high-level policy that defines routing constraints and performance objectives. 
\sys then generates P4 programs for switches in the network, which execute in a fully distributed fashion. 
Collectively, they implement a specialized version of a distance-vector protocol
that forwards traffic based on routing constraints and optimizes for the user-defined performance objectives. 
This protocol operates by generating periodic probes that traverse policy-compliant paths and collect user-defined performance metrics. 
Switches analyze the incoming probes and rank paths in real time, storing the current best next hop to reach any given destination. Since the programs run in the data plane, switches can react to performance changes quickly.
Overall, \sys is designed to achieve the following objectives: 
\begin{packeditemize}
  \item General -- operates over a wide range of policies 
  \item Reusable -- works correctly for any topology
  \item Distributed -- does not require central coordination 
  \item Responsive -- adapts to changing metrics quickly 
  \item Implementable -- on today's programmable data planes
  \item Policy-compliant -- packets only use allowed paths 
  \item Loop-free -- mitigates persistent/transient loops 
  \item Optimal -- converges to best paths under stable metrics 
  \item Stable -- mitigates oscillation under changing metrics 
  \item Efficient -- avoids undue traffic and switch overhead  
  \item Ordered -- limits out-of-order packet delivery 
\end{packeditemize}

To achieve all of these objectives, we need to address several challenges. First, to operate over arbitrary topologies, \sys requires new techniques
to search the set of possible paths for optimal routes. State-of-the-art solutions, such as Conga~\cite{alizadeh-2014-conga} and Hula~\cite{katta-2016-hula}, assume a tree-based data center topology, which makes exploring possible paths, avoiding forwarding loops, and finding optimal routes straightforward.  Second, link and path metrics can change constantly, leading to additional challenges for distance-vector protocols.
In such situations, switches may have unsynchronized views of the network in transient states, and they may make forwarding decisions based on these inconsistent views, which can result in forwarding loops and/or forwarding paths that violate the routing policy. 
Third, a na\"{i}ve solution that constantly changes routes can cause transient or even persistent chaos. 
We draw inspirations from wireless network routing~\cite{babel,aodv,perkins-1994-dsdv}, and design mechanisms that leverage programmable data planes 
to address this. 
Finally, to mitigate out-of-order packet delivery, we develop policy-aware flowlet switching~\cite{sinha-2004-flowlet} to 
forward flowlets while ensuring policy compliance. 

\para{Summary}  
We make several contributions in the design of \sys, and Figure~\ref{fig:properties} summarizes the key ideas.  
\vspace{-1mm}
\begin{packeditemize}
  \item We define a new programming abstraction that views policies as path-ranking functions, 
  and generalizes existing languages by allowing operators to specify path constraints and dynamic metrics simultaneously. 
  \item We design a new configurable, performance-aware, distance-vector routing protocol.
  \item We develop compilation algorithms that generate per-device P4 programs that implement a particular configuration of the protocol 
        based on user policy. 
  \item We have built a system prototype, and conducted thorough experiments to 
        demonstrate that \sys is competitive with state-of-the-art systems that are customized for a specific 
        topology and routing policy. 
\end{packeditemize}
\vspace{-1mm}

\para{Non-goals}  There has been abundant recent research on efficient
load-balancing strategies, especially in data centers.  The goal of this work
is not to outperform such strategies in the contexts for which they have been
manually optimized.  Rather, our goal is to facilitate the deployment of such techniques
on a much broader set of networks and with a broader collection of optimization
criteria, and to do so without asking network operators to take the time, or acquire
the expertise necessary, to write ``assembly-level'' P4 programs.

\begin{figure}[t!]
\centering
\scalebox{0.75}{
\begin{tabular}{|l|l|c|}
\hline
\textbf{Objective} & \textbf{Key idea(s)} & \textbf{Section(s)} \\
\hline

\multirow{2}{*}{General}                 & Language for performance-aware routing & \multirow{2}{*}{\ref{sec:lang}} \\ 
                                         & Policies as path-ranking functions & \\
\hline
\multirow{1}{*}{Reusable}                & Policy analyzed jointly with topology  & \multirow{1}{*}{\ref{subsec:pg}} \\
\hline
\multirow{1}{*}{Distributed}            & Synthesis of data-plane routing protocol & \multirow{3}{*}{\ref{subsec:pg}-\ref{subsec:tables}} \\ 
\multirow{1}{*}{Responsive \&}             & Periodic probes to collect path metrics & \\
\multirow{1}{*}{Implementable}           & Implemented in P4 & \\ 
\hline
\multirow{2}{*}{Policy-compliant}        & Probes and packets carry policy states & \multirow{2}{*}{\ref{subsec:pg}, \ref{subsec:tables}, \ref{subsec:probe}} \\  & Switches keep track of state transitions & \\
\hline
\multirow{3}{*}{Loop-free}               & Monotonicity analysis & \multirow{3}{*}{\ref{sec:lang}, \ref{subsec:persistentloops}, \ref{subsec:flowlet}} \\ 
                                         & Probes carry version numbers & \\ 
                                         & Early loop breaking for flowlets & \\
\hline 
\multirow{1}{*}{Optimal}              & Isotonicity analysis & \multirow{3}{*}{\ref{sec:lang}, \ref{subsec:convergence}} \\ 
\multirow{1}{*}{Stable \&}                  & Limit the frequency of probes & \\ 
\multirow{1}{*}{Efficient}               & Failure detection and metric expiration & \\
\hline
\multirow{1}{*}{Ordered}                 & Policy-aware flowlet switching & \multirow{1}{*}{\ref{subsec:flowlet}} \\ 
\hline
\end{tabular}
}
\vspace{-1.8mm}
\caption{Key ideas in \sys.} 
\vspace{-2mm}
\label{fig:properties}
\end{figure}

%% file: language.tex
\section{Policy language}\label{sec:lang}

\sys has a high-level language that can express a wide range of sophisticated policies, 
which are functions that rank network paths. 
The goal of the \sys compiler is to ensure that switches always use the best policy-compliant paths. 
The language has two main components: a) matching on paths using regular expressions, and b) computing path metrics.
As a concrete example, consider the following policy:
\begin{lang}
         minimize( if A .* then path.util else path.lat )
\end{lang}
It first classifies paths using a regular expression ($A .^*$), and then based on the
classification, it defines the rank to be either path utilization or latency. Each node will separately 
choose its best paths according to this function. 
So node A will always choose the least utilized path, 
while all other nodes will select the path with the lowest latency.

The \sys language can also capture static policies in existing systems that are not related to performance. 
For instance, FatTire~\cite{reitblatt-2013-fattire} uses regular expressions to classify legal and illegal paths (though it says nothing about the \emph{performance} of such paths). 
To route packets through a waypoint \texttt{W}, a FatTire policy would be \texttt{(.* W .*)}, which allows any path through \texttt{W} but no other paths. 
\sys can represent this by mapping all legal paths to 0 and illegal paths to $\infty$:
\begin{lang}
                     minimize( if .* W .* then 0 else $\infty$ )
\end{lang}
This policy will ensure that every node always selects a path through \texttt{W} if one exists in the network, and drops traffic otherwise; no path is preferred to a path with rank $\infty$. 

As another example, Propane~\cite{beckett-2016-propane} allows users to write policies about failover preferences. A Propane policy \texttt{(A B D) >> (A C D)} indicates a preference for sending traffic through path \texttt{A B D} and only using \texttt{A C D} if the first path is not available (\EG, a link has failed). In \sys, we can achieve the same effect by ranking paths statically as below. 
\begin{lang}
     minimize( if A B D then 0 else if A C D then 1 else $\infty$ )
\end{lang}

In \sys, it is also possible to rank paths based on multiple metrics. 
For example, suppose we prefer that A reaches D via B instead of via C, and we also prefer shorter, less utilized paths. 
This can be achieved by lexicographically ranking paths, \EG, prefer paths through B first, then shortest paths, and finally, least utilized paths.
\begin{lang}
      minimize( if A .* B .* D then (0, path.len, path.util)
                else if A .* C .* D then (1, path.len, path.util)
                else $\infty$ )
\end{lang}

\begin{figure}[t!]
  \centering
  \[ \begin{array}{rclr}
    \multicolumn{4}{l}{\textbf{Policy}} \\
    pol &::=   & \mathsf{minimize}(e)                                   & \textit{optimization} \\

    \multicolumn{4}{l}{\textbf{Expressions}} \\
    e &::= & n                                                          & \textit{constant numeric rank} \\
        &\ALT& \infty                                                   & \textit{infinite rank} \\
        &\ALT& \mathsf{path.attr}                                       & \textit{path attribute} \\
        &\ALT& e_1 \circ e_2                                            & \textit{binary operation} \\
        %&\ALT& e_1 * e_2                                                & \textit{multiplication} \\
        %&\ALT& \mathsf{max}(e_1, e_2)                                   & \textit{maximum} \\
         &\ALT& \mathsf{if}~ b ~\mathsf{then}~ e_1 ~\mathsf{else}~ e_2   & \textit{if statement} \\
       &\ALT& (e_1, \ldots, e_n)                                       & \textit{tuple} \\
    \multicolumn{4}{l}{\textbf{Boolean Tests}} \\    
    b &::=& \multicolumn{2}{l}{r \ALT e_1 \leq e_2 \ALT \mathsf{not}~ b \ALT b_1 ~\mathsf{or}~ b_2 \ALT b_1 ~\mathsf{and}~ b_2}  \\
    \multicolumn{4}{l}{\textbf{Regular Paths}} \\        
    r &::=&   \multicolumn{2}{l}{\mathsf{node\_id} \ALT . \ALT r_1 + r_2 \ALT r_1 ~ r_2 \ALT r^*} \\
  \end{array} \]
  \vspace{-4mm}
  \caption{Syntax for \sys policies.} 
  \label{fig:grammar}
\vspace{-4mm}
\end{figure}

Ranking paths using regular expressions defines strict, inviolate preferences; however, operators may have softer constraints as well: \EG, one path may be preferred up to a point, but if the utilization is too high then some traffic should be shunted
along another path instead.  To implement soft constraints, policies may make different choices based on
current path performance.
For example,
to prefer least-utilized paths when the network load is light (utilization of the path is less than 80\%),
even if those paths are long, but to prefer shortest paths when network load is heavy (and hence to save bandwidth
globally), one might use the following policy.

\begin{lang}
               minimize( if path.util < .8 
                                  then (1, 0, path.util)
                                  else (2, path.len, path.util) )
\end{lang}

Finally, to steer traffic towards or away from particular links, one may add or subtract weights.
For instance, the following policy demonstrates how to add weight to costly links AB and CD
while otherwise using simple shortest paths.  

\begin{lang}
        minimize( (if .* AB .* then 10 else 0) +
                           (if .* CD .* then 20 else 0) + path.len )
\end{lang}

Figure~\ref{fig:grammar} presents the full language syntax, and Figure~\ref{tab:policies} presents selected policy examples taken from the literature.
The key novelty of the language is that it can capture many of the \textit{static} conditions expressed by earlier work
such as FatTire~\cite{reitblatt-2013-fattire} or NetKAT~\cite{anderson-2014-netkat} as well as the \textit{relative} preferences of Propane~\cite{beckett-2016-propane}, and yet also augment such policies with \textit{dynamic} preferences based on current network conditions.

\begin{figure}[t!]
\scalebox{0.75}{
\centering
    \begin{tabular}{|l|c|}
    \hline
    \textbf{Policy} & \textbf{Implementation} \\ 
    \hline
    P1. Shortest path routing~\cite{rip}  & path.len  \\
    \hline
    P2. Minimum utilization~\cite{katta-2016-hula}  & path.util  \\
     \hline
    P3. Widest shortest paths~\cite{ma-1997-wsp}  & (path.util, path.len) \\
    \hline 
    P4. Shortest widest paths~\cite{zheng-1990-swp}  & (path.len, path.util) \\
    \hline 
    P5. Waypointing~\cite{anderson-2014-netkat} & if .*(F$_1$+F$_2$).* then path.util else $\infty$ \\ 
    \hline 
    P6. Link preference~\cite{beckett-2016-propane} & if .*XY.* then path.util else $\infty$ \\
    \hline 
    P7. Weighted link~\cite{fortz-2000-ospf} & (if .*XY.* then 10 else 0) + path.len \\ 
    \hline
    P8. Source-local preference ~\cite{alizadeh-2010-dctcp} & if X.* then path.util else path.lat \\ 
    \hline 
    P9. Congestion-aware routing~\cite{javed-2009-dst} & if path.util $<$ .8 then (1, 0, path.util)  \\ 
                                                   & else (2, path.len, path.util) \\ 
    \hline 
    \end{tabular}
}
\vspace{-3mm}
\caption{Selected \sys policies.}
\label{tab:policies}
\end{figure}

\para{Advanced policy analysis}
\sys requires policies to be \textit{isotonic} (switches have consistent preferences) and 
\textit{monotonic} (metrics do not improve for longer paths), so that switches can converge to best paths. 
If a policy is non-isotonic (\EG, P9 in Figure~\ref{tab:policies}), then the \sys compiler will decompose it into 
multiple isotonic subpolicies that can be processed separately. Due to space limitation, we refer interested readers 
to the appendix (A) for more detail. 

\para{Limitations} 
Currently, \sys does not support traffic classification, 
but extending the language with header predicates as in prior work~\cite{foster-2011-frenetic,anderson-2014-netkat}
should not present any significant intellectual challenge.
A more notable limitation involves policies that prioritize one traffic class over another.
For instance, B4~\cite{jain-2014-b4} prioritizes small, latency-sensitive user requests over
large, latency-insensitive bulk transfers.  Currently, \sys ranks paths and selects the best path for each
flowlet,
but does not compare different types of traffic in order to prefer one over the other.
We leave integration of such policies into our framework to future work.

%% file: overview2.tex
\begin{figure*}[ht!]
    \centering
    \hspace{6.5mm}
    \begin{adjustbox}{minipage=\linewidth,scale=0.85}
    \includegraphics[width=16cm]{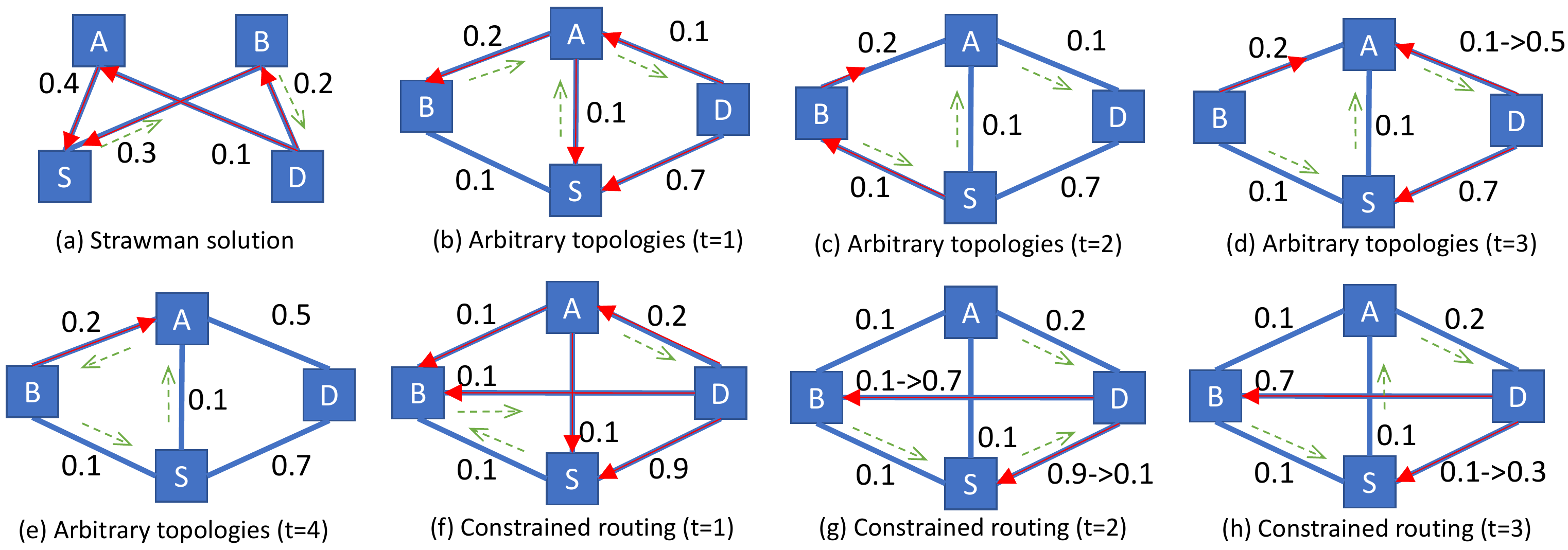}
    \caption{Supporting sophisticated policies over arbitrary topologies is challenging. (Solid, red arrows 
             represent probes, and dotted, green arrows represent packet forwarding. Links are labeled with performance metrics.)}
    \label{fig:strawman}
    \end{adjustbox}
\end{figure*}

\section{Selected Challenges} \label{sec:overview} 

The design of \sys needs to address three key challenges. To illustrate these challenges, we first describe a strawman solution that only 
works on a specific topology (data center networks) and policy (use least utilized paths). 

\para{A simple policy on a simple topology}
Consider the simple leaf-spine topology shown in Figure~\ref{fig:strawman}(a), where switch S wants to send traffic to 
switch D using the least-utilized path:
\begin{lang}
               minimize( if S.*D then path.util else $\infty$)
\end{lang}
One strawman solution 
is to use a distance-vector protocol, where each switch propagates link metrics (\emph{i.e.}, utilization) to its neighbors via 
periodic probes, and builds up a local forwarding table of ``best next hops'' to reach other switches.

More concretely, at time 1, D sends two probes to A and B carrying utilizations u(A-D)=0.1 and u(B-D)=0.2, respectively. 
Upon receiving a probe, a spine switch updates its metric, and then disseminates the probe to its downstream neighbors. 
The updated probe metric is the maximum of a) the original probe metric, and b) the utilization of the inbound link from the switch's neighbor, so the probe always carries the utilization of the bottleneck link on its traversed path. For instance, when B receives the probe 
from D, it updates the utilization to 0.3, which is the maximum of a) the original probe metric, u(B-D)=0.2, and b) the utilization 
u(S-B)=0.3; when A receives the probe from D, it updates the utilization in the probe to be 0.4, which is the maximum of 
u(A-D)=0.1 and u(S-A)=0.4. At time 2, both A and B then disseminate the updated probes to S. Now, S has received probes on both 
paths S-A-D (u=0.4) and S-B-D (u=0.3), and it chooses B as the best next hop to reach D due to its lower utilization. 
Changes in link metrics are then captured and propagated by the next round of probes.

In fact, this solution describes Hula~\cite{katta-2016-hula}, a state-of-the-art solution for utilization-aware 
routing in data centers. 

\para{Challenge \#1: Arbitrary topologies} 
\label{subsec:arbitrary-topo}
On a tree topology, simple mechanisms (\emph{e.g.}, defining a set of ``downstream'' and ``upstream'' neighbors for each switch) suffice to explore paths and prevent forwarding loops~\cite{katta-2016-hula}, but on a non-hierarchical topology, it is insufficient. 

Consider the sequence of events in Figures~\ref{fig:strawman}(b)-(e), where S prefers the least-utilized path to D. 
Suppose that at time 1, D sends out probes to A and S, and 
A propagates D's probe to B and S, with the utilizations shown in Figure~\ref{fig:strawman}(b); now, both B and S prefer to reach D via A. 
At time 2, S propagates A's probe to B about S-A-D (u=0.1), so B changes its preference to go through S; B then propagates S's probe to A (u=0.2), 
but it gets delivered only at time 4. At time 3, u(A-D) increases to 0.5, which is discovered by a new periodic probe from D to A and S. From A's perspective, the best path to reach D is still A-D, except that now the utilization is 0.5 instead. At time 4, when B's (old) probe to A arrives with u=0.2, A mistakenly thinks that 
it should instead reach D via B, not knowing that A is itself on B's best path to reach D. 
As a result, a forwarding loop S-A-B-S would form, and it will \textit{persist} as long as the link utilizations remain stable.  

In fact, it is well-known that distance-vector protocols can result in forwarding loops on an arbitrary topology. 
One might consider using a path-vector protocol instead, where probes record the paths they have traversed, 
so that switches never use paths with loops. However, since probes will be sent out frequently, carrying path information would result in much higher traffic overhead, especially in large networks. 

\para{Solution} 
Our solution is inspired by DSDV~\cite{perkins-1994-dsdv} and a more recent proposal Babel~\cite{babel}, 
which were originally developed for wireless mesh networks. 
At a high level, switches assign version numbers to probes, so that they can identify and avoid using outdated probes.
However, even with version numbers, non-monotonic policies can still create loops within a given probe period. 
So the \sys compiler additionally performs monotonicity checks on user policies. 
Finally, when \sys is integrated with flowlet switching, which pins traffic to particular paths to avoid out-of-order packet delivery, 
loops might still form when flowlet entries expire at different times. To address this, \sys lazily detects and breaks loops 
by flushing flowlet switching entries. 

\para{Challenge \#2: Constrained routing}
Supporting rich routing policies that admit way-pointing, service-chaining, or other path constraints
complicates the protocol implementation dramatically. Consider the scenario in Figure~\ref{fig:strawman}(f), where 
the policy is not only to prefer least-utilized paths, but also that traffic should never first go through B and then A due to security concerns:
\begin{lang}
               minimize(if .*BA.* then $\infty$ else path.util)
\end{lang}
Under this policy, S can only send traffic to D via a) S-D, b) S-A-D, or c) S-B-D; initially, S prefers c) (u=0.1). 
Now consider the sequence of events shown in Figures~\ref{fig:strawman}(f)-(h). 
Suppose that at time 1, the traffic from S arrives at B. At time 2, the u(B-D) increases to 0.7, and u(S-D) decreases to 0.1, 
so B updates its best next hop (to reach D) to be S, preferring the path B-S-D. At time 3, B sends the 
traffic back to S, which already forms a loop. But things can get even worse: at time 3, u(S-D) increases to 0.3, 
so S changes its preference to be S-A-D (u=0.2). So the traffic has been forwarded along a path S-B-S-A-D, which not only contains 
a loop but also violates the intended policy. 

\para{Solution}
To address this problem, \sys tags both probes and packets with policy states, which track the paths being traversed and whether these paths have satisfied the intended policy. When a switch processes a packet, it relies on 
the embedded tag to determine a local forwarding action that is compliant with the global, network-wide policy. 
When a switch changes its path preference locally, it applies a new tag on packets so that downstream switches know about the change and process the packets based on the latest preference---somewhat akin to a distributed version of consistent updates~\cite{reitblatt-2012-update}. By tagging packets at the source, different switches can then freely make independent forwarding decisions that optimize for the policy.

\para{Challenge \#3: Custom performance metrics}
Sophisticated policies may also require a more advanced probe propagation mechanism. In the mechanism we have discussed so far, a switch only propagates the probe with the best metric to its neighbors; this is due to an implicit assumption that probes arriving at a switch with worse metrics can be safely discarded, because the metrics will only degrade or remain the same as probes are propagated further along a path. However, it is only safe to discard probes when a user's policy is \emph{isotonic}~\cite{griffin-2005-metarouting}, meaning that downstream nodes respect the preference of the upstream node. Unfortunately, some useful policies are not isotonic~\cite{javed-2009-dst}. 

\para{Solution}
To address this problem, \sys first performs a static program analysis to check if a policy is isotonic. If not, 
it attempts to decompose a non-isotonic policy into multiple isotonic subpolicies. 
These different isotonic subpolicies can then be propagated separately in different probes and only recombined and evaluated later 
at the switch to make the final forwarding decision. 
To avoid sending a large number of probes, \sys uses a data structure called a product graph to minimize the number of probes while ensuring correctness. 

%% file: compilation.tex
\section{Compilation: Stable metrics}
\label{sec:compilation} 
 
\noindent  
The goal of the compiler is to generate a particular configuration of the \sys protocol that efficiently implements the desired policy 
in the data plane. 
We describe compilation in two phases.  First, in this section, we describe an algorithm that operates \textit{as if link metrics do not change}, so probes only need to be propagated once. The next section explains how this algorithm is extended to handle changing metrics. 

\begin{figure}[t!]
  \centering\includegraphics[width=5cm]{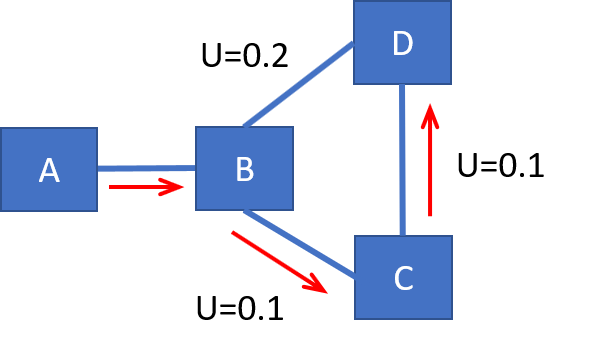}
  \caption{Na\"{i}ve solutions may lead to suboptimal paths. Node \texttt{A} uses \texttt{ABCD} even though a better path \texttt{ABD} exists.}
  \label{fig:suboptimal} 
\end{figure}

\para{Challenge}
One key challenge during compilation involves policies with conditional regular expression matches, such as
\texttt{(if r then m1 else m2)}, because nodes may rank paths differently based on the branch of the conditional they use. 
In fact, regular expressions are one source of non-isotonicity: if 
every node selects the best next hop according to its own preferences alone,
other nodes might wind up with suboptimal routes.  For example, consider
the following policy when applied to the topology in Figure~\ref{fig:suboptimal}:

\begin{lang}
               minimize( if (A B D) then 0 else path.util)
\end{lang}

\noindent 
In this example, \texttt{A} prefers path \texttt{ABD} over anything else, but 
\texttt{B} prefers the least utilized path, which is currently \texttt{BCD}. 
The correct behavior in this scenario would be for \texttt{B} to carry \texttt{A}'s traffic along path \texttt{ABD} 
while simultaneously sending its own traffic along path \texttt{BCD}. 

However, a na\"{i}ve (and erroneous) implementation may disseminate probes along the
paths \texttt{DB} and \texttt{DCB}\footnote{Recall that probes travel in the
  opposite direction to actual traffic.} and ask \texttt{B} to decide which
path is best. In this case, \texttt{B} would use the probe from \texttt{DCB} and discard the one from \texttt{DB}. 
However, if the latter probe is discarded, \texttt{A} will not receive information about its
preferred route!
To avoid this, another na\"{i}ve solution would be to propagate probes along all possible paths in the network to avoid missing good paths. 
For instance, \texttt{B} might send every probe it receives to \texttt{A}. 
However, this would lead to far too many probes, as the number of paths in a graph may be exponential in the number of nodes. 

\para{Solution}
Instead, for a conditional \texttt{(if r then m1 else m2)}, if one could determine the path with minimal metric \texttt{m1} that matches \texttt{r} using one probe, and separately determine the path with minimal metric \texttt{m2} that does not match \texttt{r} using another probe, then nodes could delay choosing their best path until both probes have been received and only then combine the information to make a decision. 
This is one concrete instance where \sys needs to decompose the non-isotonic policy (due to regular expressions) into multiple isotonic subpolicies. 
\sys achieves this by creating a compact data structure that combines all regular expressions appearing in a policy with the network topology, 
and by sending separate probes for different regular expression matches. 

\subsection{Finding policy-compliant paths}
\label{subsec:pg}

\noindent
Inspired by Merlin~\cite{soule-2014-merlin} and Propane~\cite{beckett-2016-propane}, \sys constructs a data structure called a \textit{product graph} (PG), which compactly represents all paths allowed by the policy. 

\begin{figure*}[t!]
  \centering\includegraphics[width=13cm]{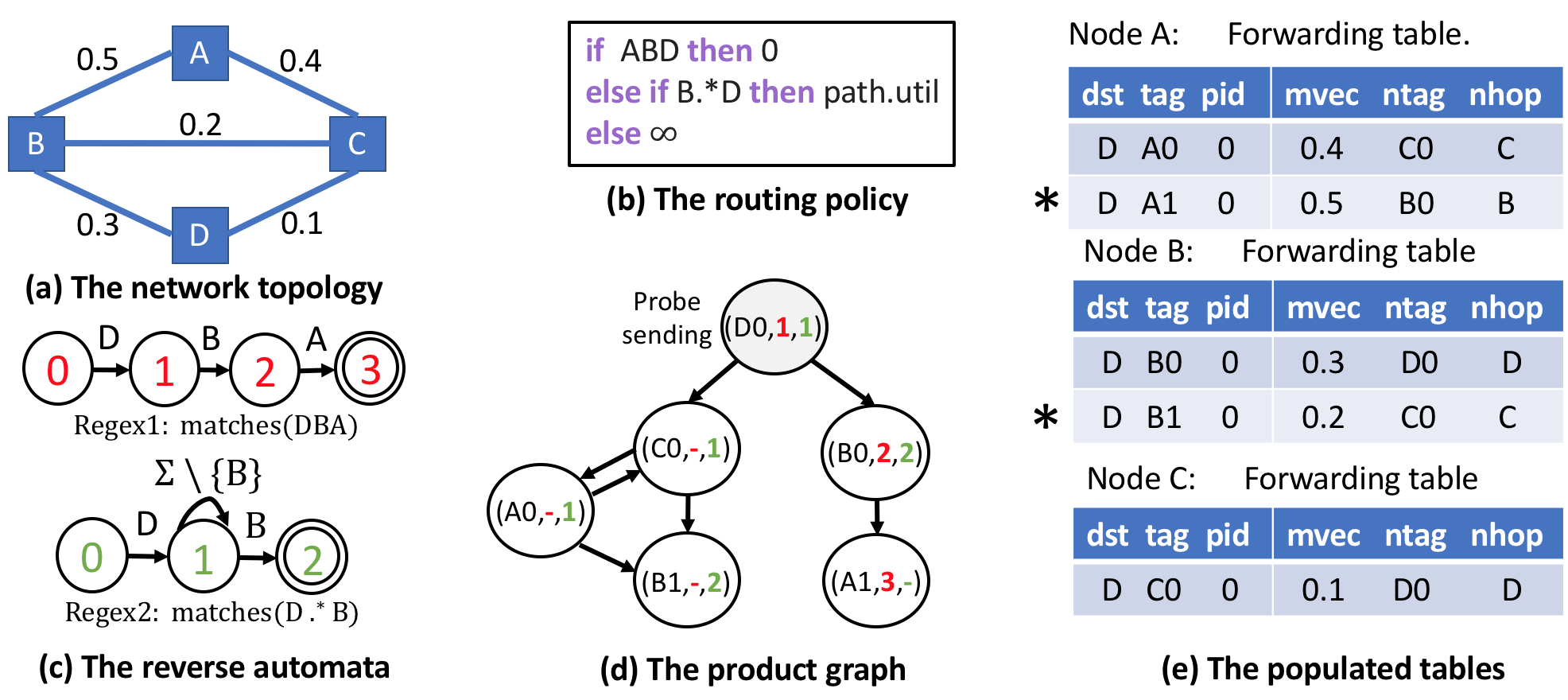}
  \vspace{-3mm}
  \caption{A running example of the compilation algorithm.}
  \label{fig:running-example}
\end{figure*}

\para{Policy automata} 
A policy's regular expressions define the different ways the shape of a path can affect its ranking. To process a policy, we first convert all such regular expressions into
finite automata.  Because probes disseminate information starting from the destination, but policies describe the direction of traffic that flows in the opposite direction, we actually construct an automaton for the reverse of each regular expression. Each automaton is a tuple $(\Sigma,Q_i,F_i,q_{0_i},\sigma_i)$. $\Sigma$ is the alphabet, where each character represents a switch ID in the network. $Q_i$ is the set of states in automaton $i$. The initial state is $q_{0_i}$. $F_i$ is the set of accepting / final states. $\sigma_i \colon Q_i \times \Sigma \rightarrow Q_i$ is the transition function.

Consider the example policy shown in Figure~\ref{fig:running-example}(b), which a) allows A to send traffic to D via the path A-B-D, b) allows B to 
send traffic to D via any path with the least utilization, and c) disallows all other paths. The \sys compiler would generate the automata shown in Figure~\ref{fig:running-example}(c). 

\para{Network topology} The construction of the automata has not considered the actual network topology, so not all automaton 
transitions are legitimate. For instance, although the automaton for \texttt{D.*B} could in principle accept a sequence 
of transitions \texttt{D-A-B}, this sequence would never happen on the network shown in Figure~\ref{fig:running-example}(a), simply because
\texttt{D} is not directly connected to \texttt{A}. Therefore, our compiler merges the topology with the automata and prunes invalid transitions. 

\para{Product graph (PG)} 
If there are $k$ automata (one for each regular expression used in the policy), then each state in the PG would have $k+1$ fields, $(X,s_1,\cdots,s_k)$, where the first field $X$ is a topology location, and $s_i$ is a state in the $i$-th automaton; there is a directed edge from $(X,s_1,\cdots,s_k)$ to $(X',s'_1,\cdots,s'_k)$, if a) $X-X'$ is a valid link on the topology, and b) for each automaton $i$, we have $\sigma_i(s_i, X') = s_i'$.

Concretely, in the PG in Figure~\ref{fig:running-example}(d), every edge represents both valid transitions on the two policy automata and a valid forwarding action on the topology. Notice, for instance, no edges exist from any \texttt{(D,*,*)} state to \texttt{(A,*,*)} state, because such edges have been eliminated due to topology constraints. As an example, there is a transition between node \texttt{D0} and \texttt{B0} in the PG because a) the topology connects \texttt{D} and \texttt{B}, and b) applying \texttt{B} to each automaton from state $1$ leads to state $2$.
Note that we use the symbol ``$-$'' to denote the special ``garbage'' state---the state
from which there is no valid transition in an automaton.

\para{Virtual nodes} 
To distinguish PG nodes from the topology locations, we call the former ``virtual nodes'' and the latter ``physical nodes''. A physical node X may have multiple virtual nodes, because probes could arrive at X via different paths, and reach different automaton states as a result. For instance, the physical node \texttt{B} has two virtual nodes \texttt{(B0,-,2)} and \texttt{(B1,2,2)}; we have labeled their location fields as \texttt{B0}, \texttt{B1} to capture this, and we call them \textit{tags}. At a high level, having multiple virtual nodes in the PG means that probes must be duplicated in order to find the best path for each path constraint. In the example, node \texttt{B} will receive two probes: one for \texttt{B0} representing a path on the way to matching regex \texttt{ABD}, and one for \texttt{B1} representing a path on the way to matching regex \texttt{B.*D}.

\para{Probe sending states} If a physical node X is a valid destination allowed by the policy (\IE, not always $\infty$), then exactly one of its virtual nodes is a \textit{probe sending} state. This state has the form $(X0,\sigma_0(q_{0_0}, X),\cdots,\sigma_k(q_{0_k}, X))$; all probes that originate from X initially carry this state. This is because, when probes start at the originating node, they can be considered to have already traversed the first hop ``X'' from the initial automata states $q_{0_i}$.

\para{Policy-compliance} 
Any path through the PG from any state to a probe sending state is a valid, policy-compliant path that is allowed by the policy. In addition, all policy-compliant physical paths also exist in the PG. 

\begin{figure*}[t]
\hrule
\vspace{1mm}
\begin{minipage}[t]{6.6cm}
\begin{algorithmic}
\footnotesize
\setcounter{ALG@line}{\value{mylineno}}
  \Function{InitProbe}{PGNode \underline{n}, ProbeId pid}
      \If {\underline{n}.isPrbSendingState}
        \State p.origin $\leftarrow$ \Call{ToTopoNode}{\underline{n}}
        \State p.pid $\leftarrow$ pid
        \State p.tag $\leftarrow$ \underline{n}.tag
        \State p.mv $\leftarrow$ \Call{InitMVec}{}
        \State \Call{MulticastProbe}{\underline{n}, p}
        \State
      \EndIf
  \EndFunction 

  \Function{MulticastProbe}{PGNode \underline{n}, Probe p}
    \State \underline{pg\_neighbors} $\leftarrow$ \Call{GetPgOutNeighbors}{\underline{n}}
    \State topo\_neighbors$\leftarrow$ \Call{ToTopoNodes}{\underline{pg\_neighbors}}
    \State \Call{Multicast}{p$\rightarrow$topo\_neighbors}
  \EndFunction
\setcounter{mylineno}{\value{ALG@line}}
\end{algorithmic}
\end{minipage}
%---------------
\hspace{-8mm}\begin{minipage}[t]{6.4cm}
\begin{algorithmic}
\footnotesize
\setcounter{ALG@line}{\value{mylineno}}

  \Function{ProcessProbe}{Switch S, Probe p}
    \State \underline{n} $\leftarrow$ \Call{NextPGNode}{S, p.tag} 
    \State p.mv $\leftarrow$ \Call{UpdateMVec}{p.inport}
    \State key $\leftarrow$ (p.origin, \underline{n}.tag, p.pid)
    \State (mv, ntag, nhop) $\leftarrow$ FwdT[key]
    \If {f(p.pid, p.mv) $<$ f(p.pid, mv)}
      \State FwdT[key] $\leftarrow$ (p.mv, p.tag, p.inport)
      \State oldKey $\leftarrow$ BestT[p.origin]
      \If {s(key) $<$ s(oldKey)}
        \State BestT[p.origin] $\leftarrow$ key
      \EndIf 
      \State p.tag $\leftarrow$ \underline{n}.tag
      \State \Call{MulticastProbe}{\underline{n}, p}
    \EndIf 
  \EndFunction

\setcounter{mylineno}{\value{ALG@line}}
\end{algorithmic}
\end{minipage}
%---------------
\hspace{-8mm}\begin{minipage}[t]{6cm}
\begin{algorithmic}
\footnotesize
\setcounter{ALG@line}{\value{mylineno}}

  \Function{SwiForwardPkt}{Packet p, Switch S}
    \State key $\leftarrow$ (p.dst, p.tag, p.pid)
    \If {fromHost(p.inport)}
      \State key $\leftarrow$ BestT[S]
      \State p.pid $\leftarrow$ key.pid
    \EndIf      
    \State (mv, ntag, nhop) $\leftarrow$ FwdT[key]
    \State p.tag $\leftarrow$ ntag
    \State \Call{SendPkt}{p, nhop}
  \EndFunction

\setcounter{mylineno}{\value{ALG@line}}
\end{algorithmic}
\end{minipage}
\hrule
\caption{Pseudocode for the synthesized per-device programs. Underlined variables are PG states.} 
\label{fig:algorithm}
\end{figure*}

\subsection{Packet forwarding}
\label{subsec:tables} 

\noindent
Before diving into the operation of the protocol itself, we first describe the structure of the forwarding (\texttt{FwdT}) tables on each switch. The compiler does not generate the actual forwarding entries for the tables---these are populated at runtime by the protocol logic based on the link metrics, which is described in the following subsection.

An entry in the forwarding table has several fields, in the form of 
\texttt{[dst$^*$,tag$^*$,pid$^*$,mv,ntag,nhop]}, 
where the fields with stars are used as keys for table lookups.
Each row of the table indicates where the given switch
will send packets destined for router \texttt{dst} when those packets
are tagged by PG node tag \texttt{tag} and probe number id \texttt{pid}.
The sender of packets will set the initial tag and the probe number
associated with the best path it has found. At each intermediate hop,
when a packet with a given \texttt{dst}, \texttt{tag}, and \texttt{pid}
matches an entry in \texttt{FwdT}, the switch will look up the next tag (\texttt{ntag})
to write into the packet to replace the current tag, and it will look
up the next hop (\texttt{nhop}) to forward the packet to.
The metrics vector (\texttt{mv}) is not used during packet forwarding,
but is used when table entries are populated (following
subsection). A property of the forwarding table is that
any \texttt{tag}-\texttt{ntag} pair found in a row of a table
should correspond to an edge in the product graph, and when
a particular \texttt{ntag} is written into a packet it is then
forwarded out the \texttt{nhop} port that leads to a topology node
corresponding to that \texttt{ntag}.  This process implies that
forwarding will always follow edges in the product graph---in other words, forwarding is guaranteed to be policy compliant so long as \texttt{ntag} and \texttt{nhop} are written consistently.

As an example, consider the \texttt{FwdT} table for switch \texttt{B}: the policy allows \texttt{B} to reach \texttt{D} either through a) \texttt{B-D}, satisfying (part of) the regular expression \texttt{ABD}, or through b) the best of \texttt{B-D}, \texttt{B-C-D}, and \texttt{B-A-C-D}, satisfying the regular expression \texttt{B.*D}. The former corresponds to the virtual node \texttt{B0} in the PG, and the latter is implemented by a combination of both \texttt{B0} and \texttt{B1}. Hence, the reader may observe that it is possible for nodes of the product graph to contribute to the implementation of more than one regular expression in the policy---this sharing improves algorithm performance as a single probe can contribute to uncovering information useful in more than one place in the policy. 

Ignoring for now how the forwarding entries were populated, consider the first entry in \texttt{B}'s table in Figure~\ref{fig:running-example}(e). The entry is generated from the virtual node \texttt{B0}: if a packet is 
at \texttt{B} with \texttt{tag=B0} and a destination \texttt{D}, then either that packet was sent from \texttt{A}, and traveled to \texttt{B} or
it was sent directly from \texttt{B}. In either case, the current best path is through the next hop \texttt{nhop=D} and it has a metric \texttt{mv=0.3}. Moreover, before 
\texttt{B} sends the packet to \texttt{D}, it should update the tag to the new virtual node's tag, which is \texttt{D0}. 
The second entry in \texttt{B}'s table is generated from \texttt{B1}. When packets are tagged with \texttt{B1}, there are two
paths they could take to \texttt{D}: \texttt{B-C-D} and \texttt{B-A-C-D}. Currently, the least utilized path is \texttt{B-C-D}, so \texttt{nhop=C} and 
\texttt{mv=0.2}. The updated tag in this case will be \texttt{C0}.
For this policy, a static analysis has determined that only one probe is needed (carrying utilization), so there is only a single probe id (\texttt{pid}) of 0.
The asterisk next to the entry for \texttt{B1} indicates that \texttt{B} prefers \texttt{B-C-D} over \texttt{B-D}, which is determined after evaluating the user policy on both paths (it is easy to evaluate a regular expression match given the PG tag since we know which regexes are accepted in each PG state). Hence, traffic sourced from \texttt{B} will choose to use the \texttt{BCD} path. Note that each source can determine its own preference: although \texttt{B} prefers \texttt{C} as the next hop, \texttt{A} will still be able to use \texttt{A-B-D} since \texttt{A}'s traffic will be forwarded using the \texttt{B0} entry.

Function \smsg{SwiForwardPkt} in Figure~\ref{fig:algorithm} summarizes the packet forwarding logic. When a packet first arrives at the switch from a host, it is treated differently. In this case, this first switch must determine the preferred path for the packet (with each path having a representative destination, PG start node and probe id), which is stored in the BestT table.

\subsection{Sending probes}
\label{subsec:probe}

\noindent
While the forwarding tables compactly encode how devices should forward traffic in a policy-compliant way, we have yet to describe how these tables are populated. To this end, the \sys compiler generates protocol logic for propagating probes from probe sending states in order to populate the tables with the best paths to each destination. 

At a high level, each node in the PG propagates probes to its neighbors. For instance, a probe starts at \texttt{D0} (\texttt{D} with tag 0) and is sent to \texttt{B0} and \texttt{C0}. \texttt{C0} updates the utilization to be 0.1 and adds this entry to its forwarding table before sending a new probe to \texttt{A0} and \texttt{B1}. Similarly, \texttt{B0} adds an entry for the probe it received from \texttt{D0} with utilization now 0.3 before sending a new probe to \texttt{A1}. \texttt{A1} receives a probe from \texttt{B0} and adds an entry with utilization 0.5, etc. \texttt{A0} receives a probe from \texttt{C0} with metric now 0.4 and adds this entry to its table before sending the probe to \texttt{C0} and \texttt{B1}. Probes will continue to propagate through the PG so long as they decrease the best available metric for that probe type and PG node. Since a static analysis ensures that policy metrics are monotonically increasing, probes will not be propagated endlessly in loops. 

To determine which entry to use for forwarding local traffic, switches compute the best path by keeping a pointer to their overall best entry (the asterisks in Figure~\ref{fig:running-example}(e)). For example, \texttt{A} must decide whether to use the entry for \texttt{A0} or \texttt{A1}. Evaluating the policy on \texttt{A0} results in $\infty$ because \texttt{A0} is not an accepting state for regex \texttt{ABD} or \texttt{B.*D}. On the other hand, evaluating the policy in \texttt{A1} results in 0 (the best rank) because \texttt{A1} is an accepting state for regex \texttt{ABD}. Hence, the asterisk appears by \texttt{A1}.

\para{Probe generation}
Probes are generated from initial PG states (\EG, \texttt{(D0,1,1)} in our example). These sending states use the procedure in \smsg{InitProbe} to initiate probes, and use \smsg{MulticastProbe} to multicast the probes along the outgoing PG edges to all downstream neighbors. Each probe carries four fields: (1) \textit{origin} denotes the topology location of the sending switch (\IE, \texttt{D} for the state \texttt{(D0,1,1)}); (2) \texttt{pid} is the probe id, as obtained from the policy decomposition; (3) \texttt{mv} denotes the metrics vector used in the policy (\IE, utilization in the example, which is initialized to a default value 0); and (4) \emph{tag} denotes the id of the PG node the probe is at.

\para{Probe dissemination} The \smsg{ProcessProbe} algorithm describes how a switch processes a probe from its neighbor. This algorithm first obtains the product graph node for the neighbor (\underline{n}). Next, it updates the metrics in the probe based on the port at which the probe arrived.  For instance, it computes and stores the maximum of the probe's current utilization and the local port's utilization. If this probe (with id $i$ and tag $t$) contains a better metric according to $f$ than what is currently associated with $i$ and $t$ in the table then it updates its forwarding table FwdT with the new next hop, next tag, and metrics vector corresponding to this probe. If an update occurs, then we also need to check if this affects the overall best choice for the switch (\IE, where the asterisk points to). The BestT table records the current best key for that choice. We look up the existing value and compare it to the current probe using the function $s$ that checks the overall value of the probe (not just per tag / probe id). Finally, the probe tag is updated to the correct value for \underline{n}, and the probe is multicast to all PG neighbors.

%% file: dynamic.tex
\section{Compilation: Unstable metrics}
\label{sec:dynamic}
 
Consider using the same solution as described in Section~\ref{sec:compilation}, but instead of sending just one probe, sending many probes periodically, one per time interval. This introduces new complications due to the lack of synchronization; certain parts of the network may be working with outdated information.
In fact, the example sequence from Section \ref{sec:overview}, Figure~\ref{fig:strawman}(b)-(f) demonstrates exactly how a problem can arise---the example culminates with the forwarding loop S-A-B-S. Notice also that in this case it is technically policy-compliant because any path from S to D is allowed, so the packet tagging mechanism would not prohibit it.

The key issue is that when switches use old probes to make decisions, loops can form. In Figure~\ref{fig:strawman}(b) the probe $p$ from B to A took a long time to propagate; by the time $p$ arrived at A, the metrics had already changed again. Concretely, $p$ was computed using an old metric u(A-D)=0.1, which had since changed to 0.5; but A still used this outdated probe and thought D was a better next hop. 

\subsection{Preventing persistent loops}
\label{subsec:persistentloops}
  
To prevent loops, we draw on ideas from Babel~\cite{babel}.  Babel avoids loops by a) distinguishing outdated probes from new ones using a version number, and b) discarding outdated probes. In our scenario, this suggests A should discard $p$ because it has an older version number, and should continue to use D as the next hop, thereby avoiding the loop. Using this scheme, when a round of probes is still in propagation, switches may have temporarily inconsistent views, so a packet may experience a transient (yet policy-compliant) loop. However, versioned probes would guarantee that persistent loops would not form~\cite{babel}. 

We note that there is a long body of work on loop prevention in routing protocols with tradeoffs being made in terms of space overhead and convergence time. \sys's compilation algorithm can potentially be integrated with different loop prevention techniques. For example one could prevent loops by adding a bit vector to each probe to record visited nodes (\IE, a path-vector protocol) at the cost of greatly increased probe overhead (one bit for every router). We opt for our approach to limit the space overhead of probes.

\vspace{-1mm}
\begin{attempt-versionedprobes}
As before, except that a) switches attach version numbers to the probes, which increase for each round; 
b) the \texttt{FwdT} table records the version number of the probe that was used to compute 
each entry; and c) before a switch updates an entry with version \texttt{v} with a probe of 
version \texttt{v$^\prime$}, it needs to check that \texttt{v$^\prime$$\ge$v}. 
\end{attempt-versionedprobes}
\vspace{-1mm}

\subsection{Probe frequency}
\label{subsec:convergence}

Versioning the probes, however, leads to an additional complexity: a node may not always be able to pick the best path. 
Consider a case where D sends probes to S every 0.2~ms along two available paths: a) $p_1$ with utilization of 0.4 
and a latency of 0.1~ms, and b) $p_2$ with utilization of 0.1 but a latency of 0.2~ms. Due to the higher latency of $p_2$, 
whenever S receives a probe from this path, it would find the probe to be outdated, since 
newer probes had arrived from $p_1$. As a result, S ends up always using $p_1$ which has a higher utilization, 
even if the policy prefers the least-utilized path $p_2$. 

We observe that this problem can be addressed by ensuring (with high probability) that old probes are fully propagated throughout 
the network before new probes are sent out. In the above scenario, if we set the probe period to be 0.2~ms or larger, then S 
would instead pick $p_2$ to be the better path after both probes have been received. 

\vspace{-1mm}
\begin{attempt-limitedprobefreq}
As before, except that the probe period needs to be larger than or equal to $0.5\times RTT$, where $RTT$ is the highest 
round-trip time between any pair of switches in the network. 
\end{attempt-limitedprobefreq}
\vspace{-1mm}

\subsection{Policy-aware flowlet switching}
\label{subsec:flowlet}
 
Since \sys can spread traffic in the same flow across multiple paths, it is important to mitigate the potential out-of-order packet delivery at the receiver side. One classic approach is \textit{flowlet switching}~\cite{sinha-2004-flowlet}, where packets in the same flow are grouped in bursts/flowlets and the same forwarding decision is applied to the entire flowlet. 
By doing so, the first packet in the flowlet is always forwarded to the best path, and subsequent packets in the same flowlet would inherit this (slightly outdated) forwarding decision. In addition to ensuring in-order delivery, this approach has the additional benefit of increasing network stability. Although each switch's best path is constantly fluctuating, at any given point, much of the current network traffic will already be pinned to a particular path to avoid out-of-order deliver. Only new flowlets will make use of the current path information.

A first attempt to implement policy-aware flowlet switching in \sys would be to have each switch maintains a table of the form \texttt{[fid$^*$,nhop,t]}, where \texttt{fid} is the flowlet ID (from hashing a packet's five tuple), \texttt{nhop} is the temporarily ``pinned'' next hop, and \texttt{t} is the timestamp of the last packet in \texttt{fid}. When the next packet in \texttt{fid} arrives, the switch computes the gap between its timestamp and \texttt{t}: if the gap is small, this packet will use the current \texttt{nhop}; otherwise, the switch expires this entry and starts a new flowlet.

\begin{figure}[t!]
\centering\includegraphics[width=7cm]{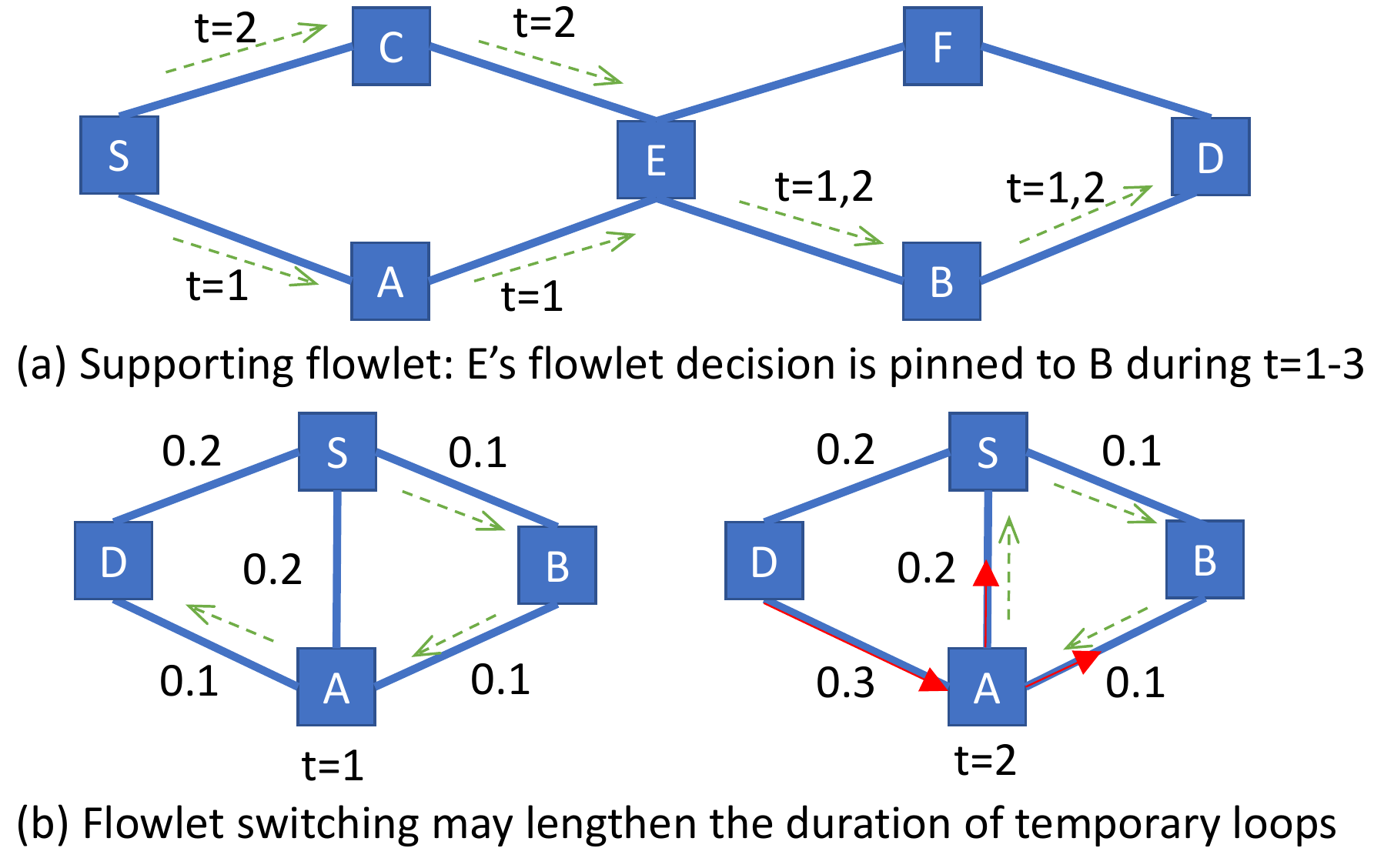}
\vspace{-4mm}
\caption{Challenges due to flowlet switching.}
\label{fig:dynamicmetrics}
\vspace{-4mm}
\end{figure}

Perhaps surprisingly, deploying such a flowlet switching implementation with \sys may result in policy violations. Consider the example in Figure~\ref{fig:dynamicmetrics}(a), where the policy prefers the least utilized of the 
upper or lower paths, but avoids the ``zigzag'' path. 

\vspace{1mm}
\begin{lang}
                if SCEFD + SAEBD then path.util else $\infty$
\end{lang}
\vspace{1mm}

Suppose that at t=1, S sends traffic to D via the lower path due to its lower utilization; using flowlet switching, 
all switches temporarily pin this flowlet to their respective next hops along the path when they receive the first packet in 
the flowlet (\emph{e.g.}, A pins to E at t=1.1, which expires at t=2.1; E pins to B at t=1.2, which expires at t=2.2; and so forth). 
At t=2, S discovers that the utilization of the upper path has improved, and changes its preference to D instead. 
However, if the packets from S arrive at E before t=2.2, which is its flowlet switching expiration time, E will continue to forward 
these packets to the lower path, causing a policy violation.

The fundamental reason for this is that flowlet switching is oblivious to any network-wide routing policy. Our solution works by making it \textit{policy-aware}. The idea is, again, that packets can carry tags that represent policy constraints, and in order to ensure policy-compliance, switches need to make forwarding decisions based on the tags. Therefore, policy-aware flowlet switching extends the table format to be \texttt{[tag$^*$,pid$^*$,fid$^*$,nhop,t]}, where \texttt{tag} and \texttt{pid} are obtained from the probe that created the forwarding entry, and \texttt{tag}, \texttt{pid}, and \texttt{fid} are match keys. This enables flowlet switching \textit{within each policy constraint and probe type}. Now, when E processes the packet at t=2.2, it would see that the packet was constrained to travel the upper path, and would use a separate flowlet switching table entry for the upper path to forward it.

\vspace{-1mm}
\begin{attempt-flowlet}
As before, except that switches perform \textit{policy-aware} flowlet switching by maintaining multiple entries for the same flowlet, 
each for a different path constraint/tag and probe type. 
\end{attempt-flowlet}
\vspace{-1mm}

\subsection{Handling failures}
Switches also need to discover new best paths when links or switches fail. Suppose that the best path for S to reach D is S-A-D, but the link A-D goes down at some point. We need to ensure that S will learn about the failure and change to another available path if one exists. Our solution is to first detect failed links, and then to expire flowlet entries when their next hop is along a link that is believed to be failed.

\vspace{-1mm}
\begin{attempt-expiration}
As before, except that, and a flowlet entry is expired when a packet arrives at a switch and is going to be forwarded by the flowlet entry, and the next hop is along a failed link.  
\end{attempt-expiration} 
\vspace{-1mm}

This approach ensures that the switch will have to route around the failure in the future. Note that as long as there is a sound way to detect failed links, this scheme will work. In our implementation of \sys, we have a switch mark a link as failed when there have been no probes along the link for $k$ probe periods, where \texttt{k} is a parameter that determines how fast (in terms of RTTs) failures should be discovered. However, other methods are possible as well, for example if the hardware could locally detect the failure of a link.

\subsection{Breaking transient loops} 
\label{subsec:temporaryloops} 

As discussed in Section~\ref{subsec:persistentloops}, even with versioned probes, transient loops may still occur when probes are in propagation. 
Figure~\ref{fig:dynamicmetrics}(b) is a concrete example. At t=1, the best path for S to reach D is S-B-A-D. Then, at t=2, A receives a 
probe from D carrying a worse metric, therefore it propagates the probe to S and B. Before this probe arrives at S and B, A learns of the better path through S, and traffic that is 
already in flight will be forwarded along a transient loop S-B-A-S; this loop will be broken once S and B receive the new probe because it 
has a higher version number. 

Interestingly, flowlet switching may lengthen the duration of transient loops because flowlet switching decisions may expire at 
different times across hops. Suppose that A's timer expires at t=3, and it starts using the new best next hop S to reach D; 
however, the timers at S and B do not expire util t=4. Then the traffic would continue to be forwarded in the loop S-B-A-S 
regardless of the newer probe, until S and B have updated their flowlet switching decisions.

We address this by detecting loops lazily and flushing the offending flowlet switching entries upon detection. 
Concretely, each switch maintains a \textit{loop detection} table 
\texttt{\{pkt\_hash$^*$,maxttl,minttl\}}, which maps a packet's CRC hash to the maximum and minimum TTL 
values seen at this switch. \texttt{$\delta$=maxttl-minttl} should be stable in the absence of loops: it is the difference 
between the longest and the shortest paths packets could have traversed to reach the current switch. 
However, when there is a loop, $\delta$ would continue to grow. Therefore, switch detects a potential loop (with false positives) 
when its $\delta$ exceeds a threshold. When this happens, the switch expires its flowlet switching 
decision, and starts a new flowlet using the latest metric in the \texttt{FwdT} table. Hence, we arrive at our final solution below.

\begin{final}
As before, except that switches use loop detection tables to detect and break loops by refreshing 
their flowlet switching decisions using the latest metrics. 
\end{final}

%% file: evaluation.tex
\section{Evaluation}
\label{sec:eval}

\noindent 
We aim to answer three main questions in our evaluation: 
a) How well does \sys scale to large networks? b) How competitive is \sys compared to hand-crafted systems? 
and c) How well does \sys work on general topologies? 

\subsection{Prototype implementation and setup}
\label{subsec:setup}

\noindent 
Our \sys prototype consists of $7485$ lines of code in F\#~\cite{fsharp}. It processes
policy and topology descriptions, and then generates device-local P4 programs.  In addition to implementing the algorithms
described in this paper, it also performs 
a variety of optimizations, such as minimizing the number of tags,
minimizing the forwarding table sizes, and 
reducing the number of bits to represent the tags. 

\para{Experimental setup} 
Our experiments were performed on a Dell OptiPlex 7060 computer, with an Intel i7-8700 CPU with 6 cores and
12 hyperthreads at 3.2~GHz, 16~GB of RAM, and a 64-bit Ubuntu 16.04 OS.
We have validated our prototype 
both in Mininet~\cite{web-mininet} and on ns-3; but our main results were obtained from ns-3 since 
Mininet does not support large topologies as efficiently.
We have used a custom tool that can compile P4 programs to run on ns-3. 

We have used three types of network topologies: a) data center 
topologies, b) random graph topologies, and c) real-world topologies (e.g., the Abilene network~\cite{abilene} and those 
from Topology Zoo~\cite{topology-zoo}). 
Our baseline systems for data center networks are ECMP and Hula~\cite{katta-2016-hula}, both of which are specifically designed 
for a Fattree topology. ECMP balances traffic randomly without considering network load, and Hula always chooses the least-utilized path among all shortest paths. 
Our baseline system for arbitrary graphs is
SPAIN~\cite{mudigonda-2010-spain}, which statically (\IE,
independently of network load) selects multiple
paths
along which to route flows.  
We used two workloads obtained from production networks for our evaluation: 
a web search workload~\cite{alizadeh-2010-dctcp}, and a cache workload~\cite{roy-2015-social}.
Due to space constraints, we have included a subset of the results as appendix (B-E). 

\subsection{Compiler scalability}
\label{subsec:scalability}

\begin{figure}[t!]
	\centering
	\begin{subfigure}[b]{4.1cm}
		\centering
		\includegraphics[width=4.1cm]{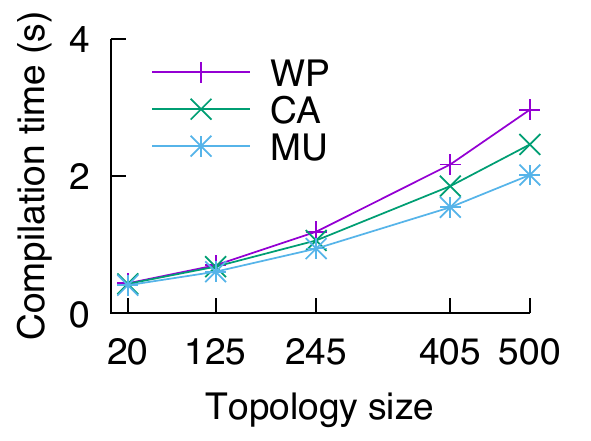}
		\vspace{-3mm}
		\caption{Fat-tree topologies}
		\label{fig:scalability-tree}
	\end{subfigure}
	~
	\begin{subfigure}[b]{4.1cm}
		\centering
		\includegraphics[width=4.1cm]{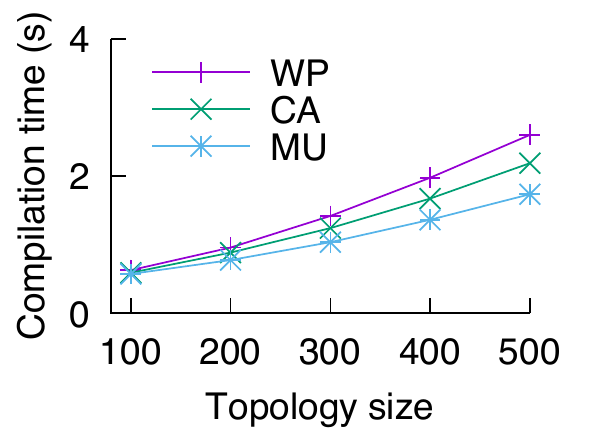}
		\vspace{-3mm}
		\caption{Random networks}
		\label{fig:scalability-random}
	\end{subfigure}
	\caption{The \sys compiler scales well to large network sizes and sophisticated policies (unit: seconds).}
	\label{fig:compiler-scalability}
\end{figure}

\noindent 
To test the scalability of our compiler, we used topologies of varying sizes 
from $20$ to $500$ nodes. For each topology, we evaluated three different policies: 
a) minimum utilization (MU: no regular expressions, single performance metric), b) waypointing (WP: three regular expressions, 
single performance metric), and c) congestion-aware routing (CA: no regular expression, non-isotonic policy with 
two performance metrics). 
Figure~\ref{fig:compiler-scalability} presents the results. 
The compiler scales roughly linearly with 
topology size, and completes in seconds on topologies with hundreds
of nodes.  Use of regular expressions increases product graph size
and hence compilation time.  In addition, non-isotonic policies
add some small amount of overhead due to the additional policy analysis. 

\begin{figure}[t!]
	\centering
	\begin{subfigure}[b]{4.1cm}
		\centering
		\includegraphics[width=4.1cm]{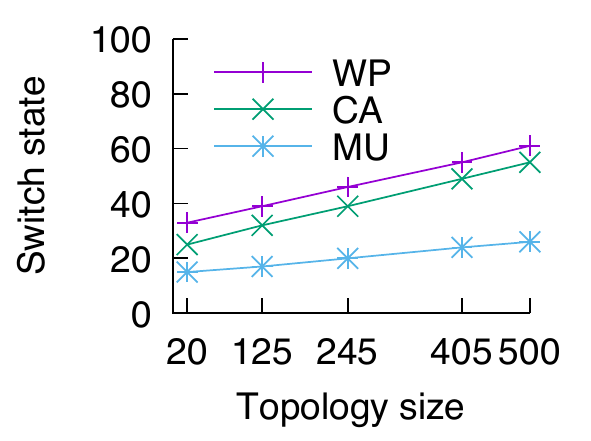}
		\vspace{-3mm}
		\caption{Fat-tree topologies}
		\label{fig:space-tree}
	\end{subfigure}
	~
	\begin{subfigure}[b]{4.1cm}
		\centering
		\includegraphics[width=4.1cm]{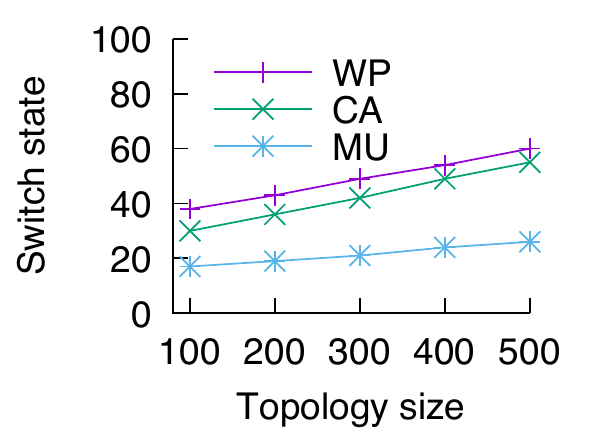}
		\vspace{-3mm}
		\caption{Random networks}
		\label{fig:space-random}
	\end{subfigure}
	\caption{The \sys compiler generates programs with low memory overhead (unit: kB).}
	\label{fig:compiler-space}
\end{figure}

Figure~\ref{fig:compiler-space} further plots the switch state used by the generated P4 programs. 
As expected, WP and CA require 
more state than MU: WP's regular policy requires tag processing to track automaton states, 
and CA's non-isotonic policy requires a separate table for each metric in the decomposed policy (\IE, separate entries for different \texttt{pid} values).
However, no more than $70$~kB of switch state was necessary in any experiment---a tiny fraction of
the available space on modern switch hardware (tens of megabytes)~\cite{tofino}.

\subsection{Performance: Data center topology}
\label{subsec:performance-tree}

\begin{figure}[t!]
	\centering
	\begin{subfigure}[b]{4.1cm}
		\centering
		\includegraphics[width=4.1cm]{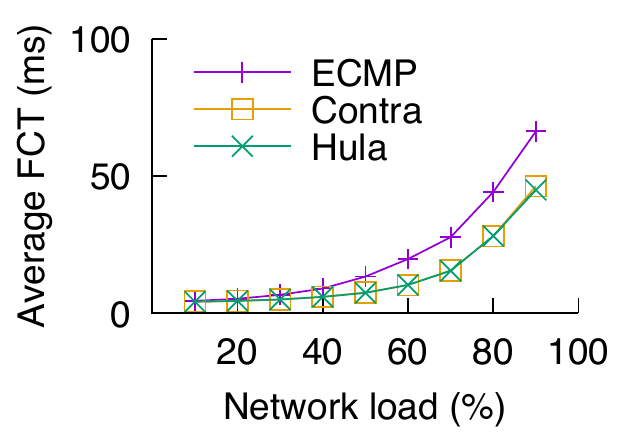}
		\vspace{-3mm}
		\caption{The web search workload}
		\label{fig:fct-ws-tree}
	\end{subfigure}
	~
	\begin{subfigure}[b]{4.1cm}
		\centering
		\includegraphics[width=4.1cm]{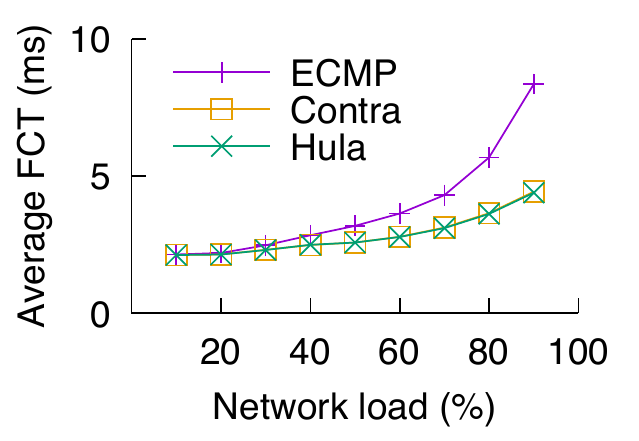}
		\vspace{-3mm}
		\caption{The cache workload}
		\label{fig:fct-cache-tree}
	\end{subfigure}
	\caption{\sys achieves a similar FCT as Hula, outperforming ECMP considerably.}
	\label{fig:fct-bothworkloads-tree}
\end{figure}

\noindent 
Our performance evaluation starts with the simplest case: a data center network topology. 
We compare \sys with ECMP and Hula in terms of their flow completion time (FCT), 
and note that the latter two mechanisms are designed specifically for
a Fattree topology, whereas \sys can work over any topology. 

In our topology, we used $32$ hosts with $10$~Gbps links, 
a bisection bandwidth of $40$~Gbps, and an oversubscription ratio of 4:1. 
Half of these hosts were configured as senders, and the other half receivers. 
We set the probe period to $256\mu$s for both \sys and Hula, and the flowlet timeout to be $200\mu$s for 
all systems. All links have a buffer size of $1000$~MSS by default. 
Moreover, we tuned the desired network load from 10\% to 90\% by adjusting the flow arrival times, 
and obtained the FCT for each setting. 

\para{Symmetric Fattrees} 
Figure~\ref{fig:fct-bothworkloads-tree} shows our results for the datacenter setting. As we can see, both \sys and Hula outperform ECMP considerably 
because they balance traffic based on network load. At 90\% load, they reduce the average FCT by 30\% for web search dataset and by 47\% for cache dataset. 
Hula outperforms \sys slightly, by $0.33\%$ on average across different datasets and network loads. 
This is because Hula knows statically what paths are shortest paths (and hence what ports to send probes from), whereas \sys has to discover this information dynamically (\IE, by carrying the path length as well as the utilization, and also by sending probes both ``up'' and ``down'' at each level in the datacenter)---hence \sys sends more probes than Hula in order to achieve generality over different topologies and policies. Further compiler optimizations could likely reduce this gap further (\EG, by identifying shortest paths statically).

\begin{figure}[t!]
	\centering
	\begin{subfigure}[b]{4.1cm}
		\centering
		\includegraphics[width=4.1cm]{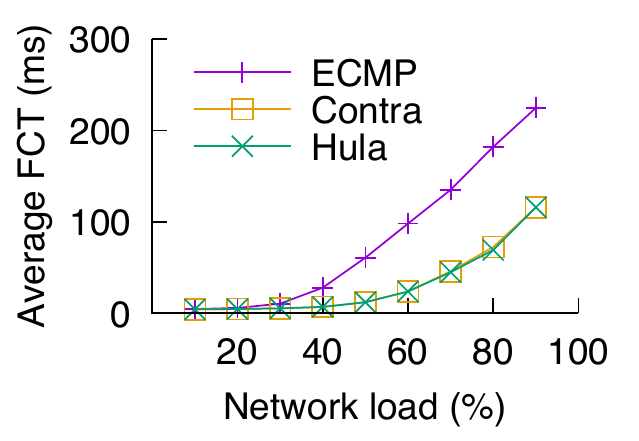}
		\vspace{-3mm}
		\caption{The web search workload}
		\label{fig:fct-ws-tree-asymmetric}
	\end{subfigure}
	~
	\begin{subfigure}[b]{4.1cm}
		\centering
		\includegraphics[width=4.1cm]{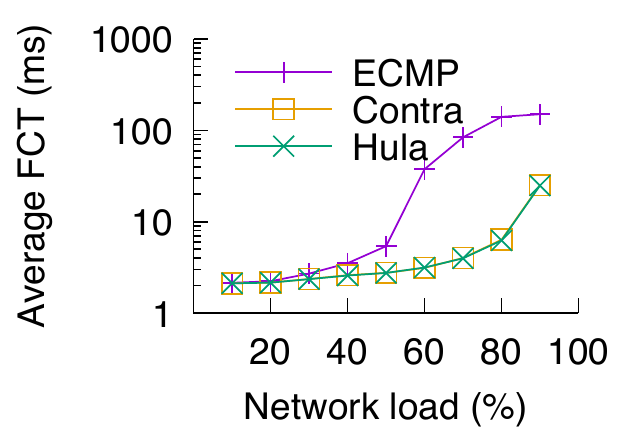}
		\vspace{-3mm}
		\caption{The cache workload}
		\label{fig:fct-cache-tree-asymmetric}
	\end{subfigure}
	\caption{\sys achieves a significantly shorter FCT on an asymmetric topology with a failed link.} 
	\label{fig:fct-bothworkloads-tree-asymmetric}
\end{figure}

\begin{figure}[t!]
	\centering
	\includegraphics[width=7cm]{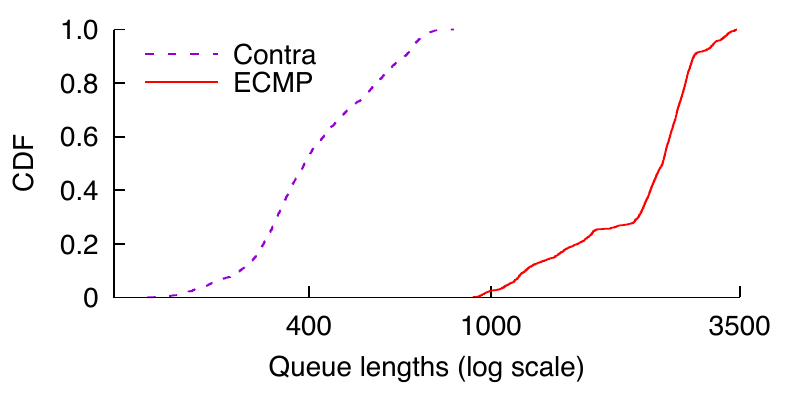}
	\caption{\sys has shorter queues than ECMP.} 
	\label{fig:queue-ws-asym-tree-60}
\end{figure}

\para{Asymmetric Fattrees} 
Next, we ran the same experiment after injecting a failure on a link between an aggregation switch and a core switch, 
so that the topology became asymmetric. Figure~\ref{fig:fct-bothworkloads-tree-asymmetric} shows the FCT for this setting. 
In this case, we found that ECMP incurred heavy traffic loss beyond 50\% network load, even though 75\% of 
all capacity remains after the link failure. The average FCT was inflated by $3.18\times$ for web search dataset and $8.72\times$ for cache dataset.
In contrast, \sys and Hula only had an increase of $1.80\times$ for web search dataset and $1.67\times$ for cache dataset, 
relative to the FCTs on the symmetric topology.

We further measured the queue sizes under ECMP and \sys with 60\% workload on the web search dataset. 
Figure~\ref{fig:queue-ws-asym-tree-60} shows the results. %CDF of the queue lengths. 
We found that \sys's queue lengths never exceeded 
$1000$~MSS, whereas ECMP had lengths larger than this value more than 97\% of the time, 
which caused heavy traffic loss when the queues are full. 

\begin{figure}[t!]
	\centering
	\includegraphics[width=7cm]{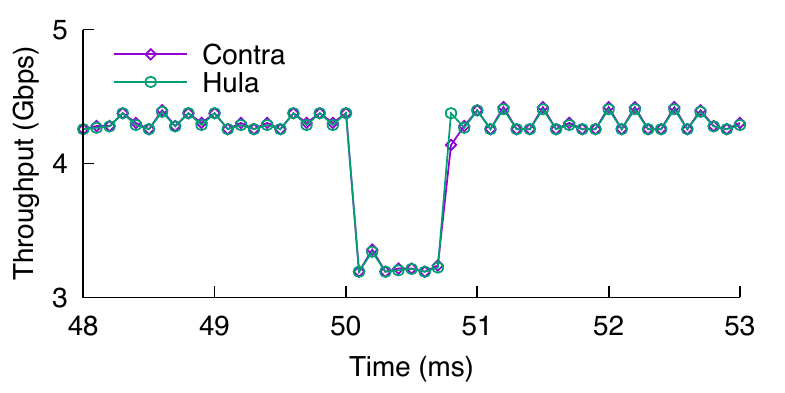}
	\caption{\sys recovers from the link failure within $1$~ms.} 
	\label{fig:recovery}
\end{figure}

We also tested the time for \sys to respond to link failures. 
Figure~\ref{fig:recovery} shows the aggregate throughput before and after a link failure, using UDP 
workloads at a stable rate of $4.25$~Gbps. We brought down an aggregate-core link at 
$t=50$~ms. \sys successfully detected this failure $800\,\mu$s afterwards, which is close to 
the failure detection threshold ($3\times$RTT=$768\mu$s) that we used for this experiment. 
Upon detection, \sys routed around the failure and was able to recover the throughput within $1$~ms. 
We have found Hula to perform similarly to \sys, as shown in the same figure. 

\subsection{Performance: Arbitrary topologies}
\label{subsec:performance-abilene}

\noindent 
We now turn to evaluate the performance of \sys on general topologies. 
We modeled our network after the Abilene~\cite{abilene} topology, 
configured all links to be $40$~Gbps, and randomly chose 
four pairs of senders/receivers. Since Hula is specialized to a Fattree topology and will not work outside of this context, and since ECMP will not load balance when there is only a single shortest path, we have used two 
other baselines: a) shortest path routing (SP), which simply sends traffic to the shortest paths, and b) SPAIN~\cite{mudigonda-2010-spain}, 
which precomputes all paths using (static) heuristics that avoid
overlap, and then load balances between these paths.

\begin{figure}[ht!]
	\centering
	\begin{subfigure}[b]{4.1cm}
		\centering
		\includegraphics[width=4.1cm]{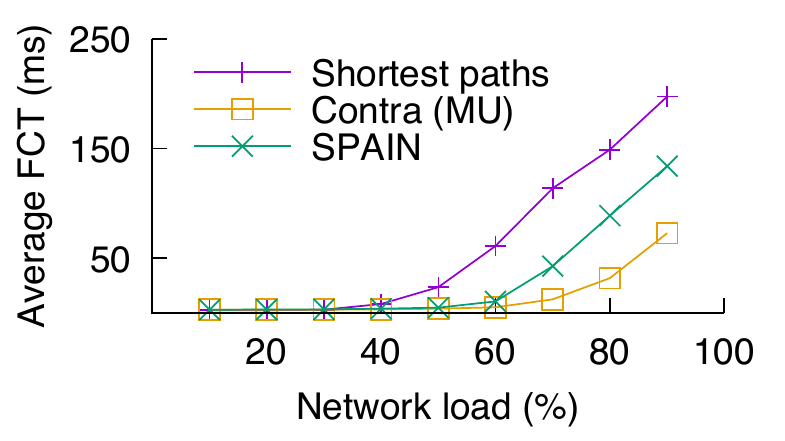}
		\vspace{-3mm}
		\caption{The web search workload}
		\label{fig:fct-ws-abilene}
	\end{subfigure}
	~
	\begin{subfigure}[b]{4.1cm}
		\centering
		\includegraphics[width=4.1cm]{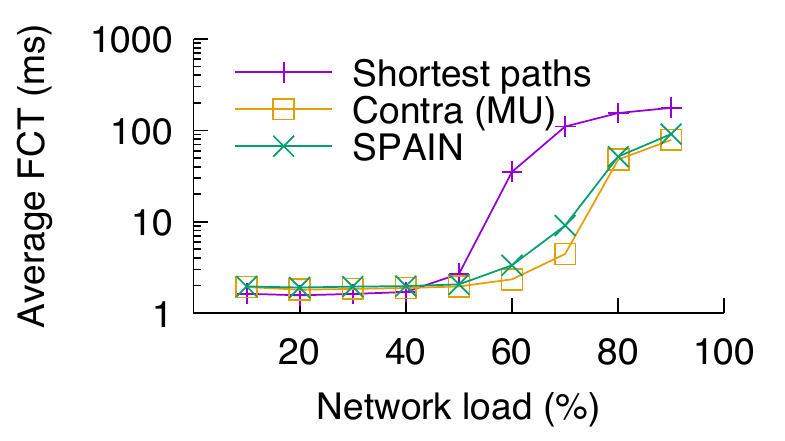}
		\vspace{-3mm}
		\caption{The cache workload}
		\label{fig:fct-ca-abilene}
	\end{subfigure}
	\caption{\sys outperforms SPAIN in FCT.}
	\label{fig:fct-abilene}
\end{figure}

Figure~\ref{fig:fct-abilene} shows the FCT for these different systems. A na\"{i}ve strategy that simply chooses 
shortest paths performs the worst. Since SPAIN can utilize multipath routing, it outperforms SP by 32.5\% 
on average for the web search workload and 26.9\% on for the cache workload. 
\sys achieves the best performance among the three: it evenly distributes traffic based on path utilization, 
and reduces FCT relative to SPAIN by 31.3\% on average for the web search workload and 13.8\% for the cache workload. 

\subsection{Protocol overhead}

\begin{figure}[t!]
	\centering
	\includegraphics[width=7cm]{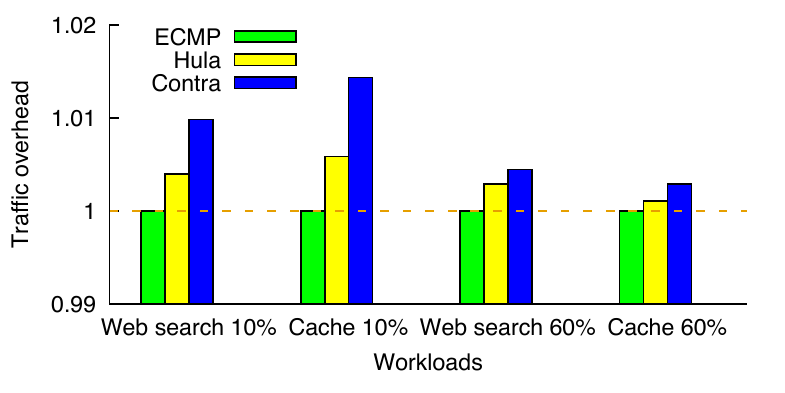}
	\vspace{-2mm}
	\caption{The traffic overhead of \sys is low.} 
        \vspace{-2mm}
	\label{fig:traffic-overhead-tree}
\end{figure}

To evaluate the traffic overhead incurred by \sys due to packet tags and probes, we measured the amount of traffic 
sent over the network by \sys, Hula, and ECMP at 10\% and 60\% network load. 
Figure~\ref{fig:traffic-overhead-tree} shows the traffic overhead as normalized by ECMP as the baseline. 
Across workloads, \sys incurred 0.79\% more traffic than ECMP, and 0.44\% more than Hula, 
which seems to be reasonable. We have similar observations for \sys on the Abilene network, as well as for 
the WP policy, and we have included these results in the appendix (D+E). 

Another type of overhead comes from transient loops, which may arise
as performance metrics change and nodes are temporarily out of sync. 
To quantify this, we measured the amount of traffic 
that has experienced transient loops using the MU policy on a Fattree and on Abilene with 60\% workload. 
We found that 0.026\% and 0.007\% of the traffic traveled in 
a loop, respectively, and that our loop detection mechanism 
successfully broke such loops upon detection. 

%% file: relatedwork.tex
\section{Related Work}
\label{sec:related}

\para{Traffic engineering} 
Centralized traffic engineering solutions such as 
B4~\cite{jain-2014-b4} and SWAN~\cite{hong-2013-swan} perform load balancing for wide-area networks, 
and Hedera~\cite{alfares-2010-hedera} and MicroTE~\cite{benson-2011-microte} for data centers.
Distributed traffic engineering solutions like TeXCP~\cite{kandula-2005-texcp} and MATE~\cite{elwalid-2001-mate} perform load balancing 
across ingress-egress paths in wide-area networks; Halo~\cite{michael-2015-halo} 
performs load-sensitive routing by solving an optimization problem in router software.
Since these solutions typically involve router software or centralized controllers when adjusting to traffic load, adaptations 
only happen at a much coarser time granularity than \sys. 

\para{Data-plane load balancing}  
Recent work on data-plane load-balancing mechanisms, such as Hula~\cite{katta-2016-hula}, Conga~\cite{alizadeh-2014-conga}, and 
DRILL~\cite{ghorbani-2017-drill} perform load balancing at a finer granularity and achieve a faster response to changes in network 
load. They are also utilization-aware---an improvement over simpler mechanisms such 
as ECMP, which splits traffic randomly regardless of network conditions. 
However, these are mostly point solutions that are specialized for a
particular topology with a hard-coded
policy. \sys supports a wide range of policies, and works over arbitrary topologies.

\para{Routing protocols and route updates} 
There is a long line of work on distance-vector routing protocols with a variety of loop prevention 
techniques~\cite{garcia-1993-diffusing,murthy-1998-loopfree,albrightson-1994-eigrp,perkins-1994-dsdv,babel,aodv} with different tradeoffs 
between overhead, convergence time, (in)stability, etc. \sys is most related to 
DSDV~\cite{perkins-1994-dsdv}, AODV~\cite{aodv}, and Babel~\cite{babel}, which use sequence numbers on route updates to achieve timely 
convergence.
Compared to existing work, the novelty of \sys lies in its use of programmable data planes to implement a wide array of distance-vector protocols in the presence of
unstable metrics, and its design of policy-aware flowlet switching mechanisms. 

\para{Regular languages for networking} 
NetKAT~\cite{anderson-2014-netkat}, Merlin~\cite{soule-2014-merlin}, FatTire~\cite{reitblatt-2013-fattire}, 
path queries~\cite{narayana-2016-pathqueries}, and Propane~\cite{beckett-2016-propane} all use regular expressions, like \sys, to 
specifying path constraints.
A key difference is that \sys supports specification and implementation of route preferences based on dynamic network conditions. 

%% file: conclusion.tex
\section{Conclusion}
\label{sec:conclusion}

\noindent 
We have presented \sys, a system for specifying and enforcing performance-aware routing policies. 
Policies in \sys are written in a declarative language, and compiled to switch programs that 
run on the data plane to implement a variant of distance-vector protocols. 
These programs generate probes to collect path metrics, and
dynamically choose the best paths along which
to forward traffic. Our evaluation shows that the compiler scales well to large topologies, 
and that the synthesized switch programs can achieve performance competitive with
hand-crafted solutions that are specialized to particular topologies and
hard-coded policies.
However, it is also substantially more general, allowing network
operators to specify a wide range of policies and to apply these
policies
to networks with arbitrary topologies.